\documentclass[prl,superscriptaddress,amsmath,amssymb,showpacs,noshowkeys,a4paper,twocolumn]{revtex4-1}

\usepackage{graphicx}
\usepackage{dcolumn}
\usepackage{bm}
\usepackage[usenames]{color}
\usepackage{epstopdf}
\definecolor{mygray}{gray}{0.5}
\usepackage{enumerate}
\usepackage [babel=true]{csquotes}

\newcommand{\affmsc}{\affiliation{Mati\`ere et Syst\`emes Complexes, CNRS and Universit\'e Paris Diderot UMR 7057, B\^atiment Condorcet, 10 rue Alice Domon et L\'eonie Duquet, 75013 Paris, France}}

\newcommand{\afflpmcn}{\affiliation{Institut Lumi\`ere Mati\`ere, CNRS, Universit\'e Lyon 1, UMR 5306, Universit\'e de Lyon 69622 Villeurbanne cedex, France}}

\begin{document}

\title{Orbits and reversals of a drop rolling inside a horizontal circular hydraulic jump}

\author{A.~Duchesne}
\affmsc

\author{C.~Savaro}
\affmsc

\author{L.~Lebon}
\affmsc

\author{C.~Pirat}
\afflpmcn

\author{L.~Limat}
\affmsc

\begin{abstract}
We explore the complex dynamics of a non-coalescing drop of moderate size inside a circular hydraulic jump of the same liquid formed on a horizontal disk. In this situation the drop is moving along the jump and one observes two different motions: a periodic one (it orbitates at constant speed) and an irregular one involving reversals of the orbital motion. Modeling the drop as a rigid sphere exchanging friction with liquid across a thin film of air, we recover the orbital motion and the internal rotation of the drop. This internal rotation is experimentally observed.

\end{abstract}

\maketitle 
	Usually when a drop comes into contact with the same liquid or a solid surface it coalesces with liquid or spreads on the solid. There are however exceptions, which could be called situations of \enquote{non wetting}, when a very thin layer of air or vapor remains trapped between the drop and the substrate such as in the well known Leidenfrost effect \cite{Biance_Leidenfrost}. In these levitation situations, the disappearance or reduction of friction with the substrate leads to remarkable dynamics of fluids with (nearly) no contact: unusual shapes (Poincar\'e's shapes) of drops rolling down a plane \cite{Aussillous_Nature}, particle-wave duality of drops bouncing on a vibrated bath \cite {from_bouncing_to_floating, JFMSuzie, Couder_Diffraction}, drop motions induced by its own harmonic modes on vibrated viscous bath \cite{NJOPRoller, Dorbolo_08},  chaotic behavior of a droplet on a soap film \cite{Gilet_trampoline_PRL, Gilet_trampoline_JFM}, etc.\\
	\indent Another case of mobile drops in a \enquote{non-wetting} situation has been reported by Sreenivas \textit{et al.} \cite{Sreenivas_JFM} and Pirat \textit{et al.} \cite{Pirat_gyroscopic_droplet}, when a drop is deposited inside a circular hydraulic jump of the same liquid \cite{Watson_JFM, Bohr_1993, Bohr_96, Bush_2003}. A thin layer of air  is entrained underneath the drop by the supercritical flow of liquid feeding the jump, that prevents coalescence, the drop remaining trapped at the shock front with a strong internal rotation. In a previous letter \cite{Pirat_gyroscopic_droplet} we have shown that this \enquote{non wetting} situation was also associated to remarkable dynamics in the case of a slightly inclined jump. In a well defined range of flow rate, a drop of moderate size (typically close to the capillary length) undergoes a gyroscopic instability, with surprising motions along the jump perimeter, leading to oscillations around the lowest equilibrium position. \\
	\indent In the present Letter, the authors investigate experimentally a new situation: the drop is now deposited inside a jump, formed on a perfectly horizontal circular disk, taking care to have uniform boundary conditions at large scale. Because of the disappearance of any reference equilibrium position, the drop is in fact always moving around the jump, with at least two different possible states: clockwise or anti clockwise regular orbital motion and a more complex state in which the drop \enquote{hesitates} between these two possible motions, leading to an irregular behavior involving complex mechanisms of reversal. These reversals of motions are reminiscent of those observed in more complex hydrodynamical systems \cite{RBconvection_reversal_Lohse_PRL, Magnetic_reversal_Berhanu_2007_EPL, 3_reversal_experiments_Fauve_12}, such as Rayleigh-B\'enard convection, Kolmogorov flow or the \enquote{dynamo} instability. They can also be observed in a meteorological context (quasi-biennial oscillations of high altitude winds)  \cite{QBO_Am_Geo}.  These phenomena are attracting presently a great interest from a large community ranging from hydrodynamics and non-linear physics to geophysics and meteorology. Our system constitutes perhaps the simplest experiment that one can build to observe these reversals in fluid dynamics, and in particular without turbulence.\\
	\begin{figure}[t]
\begin{centering}
\includegraphics[width=0.6 \columnwidth]{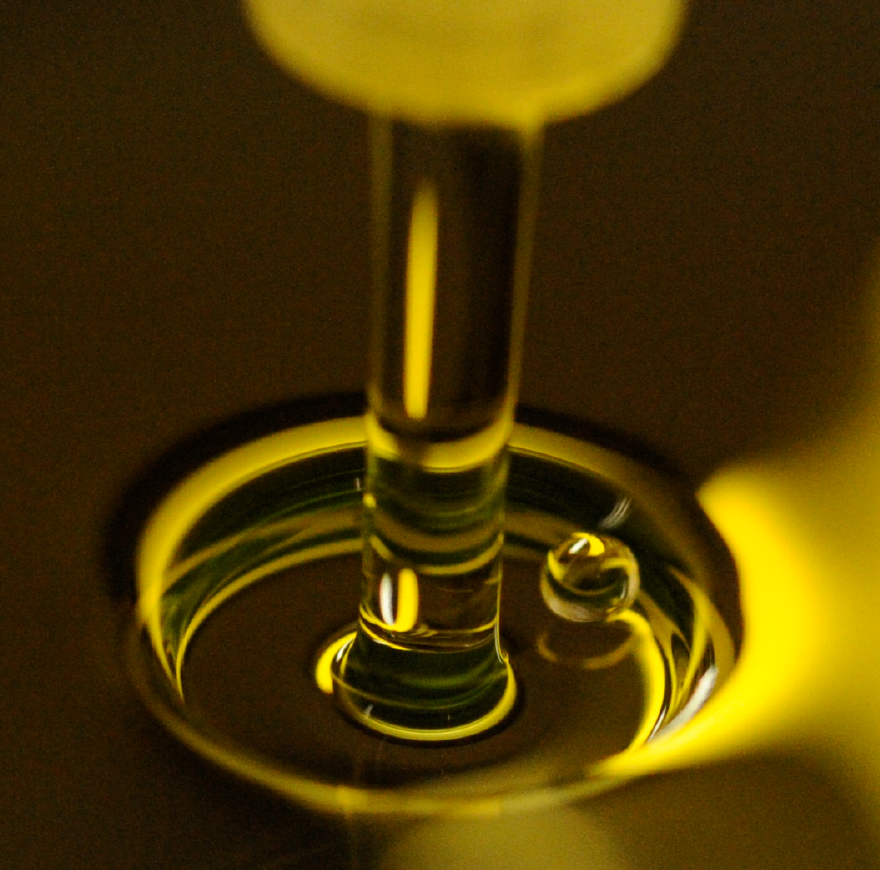}
\caption{\label{fig1} A drop of radius $a=1$ mm of silicone oil in non coalescence state inside a horizontal hydraulic jump of the same liquid ($R=5$ mm).}
\end{centering}
\end{figure}
	\indent A model of this new situation is proposed and provides a reasonable description of the regular orbital motion. As this model involves an internal rotation of the drop that has never been characterized, experiments were developed and allow us to observe and study this internal rotation. As predicted by our model, this rotation exists, but its dependence upon drop radius is more complex than expected from an analogy with a rigid sphere supported by an air film under Couette flow, suggesting that drop shape distortions (as well a jump shape distortions) should be considered to get a better description.\\
	\begin{figure}[t]
	    \centering
    \begin{tabular}{cc}
      \includegraphics[width=0.8 \columnwidth]{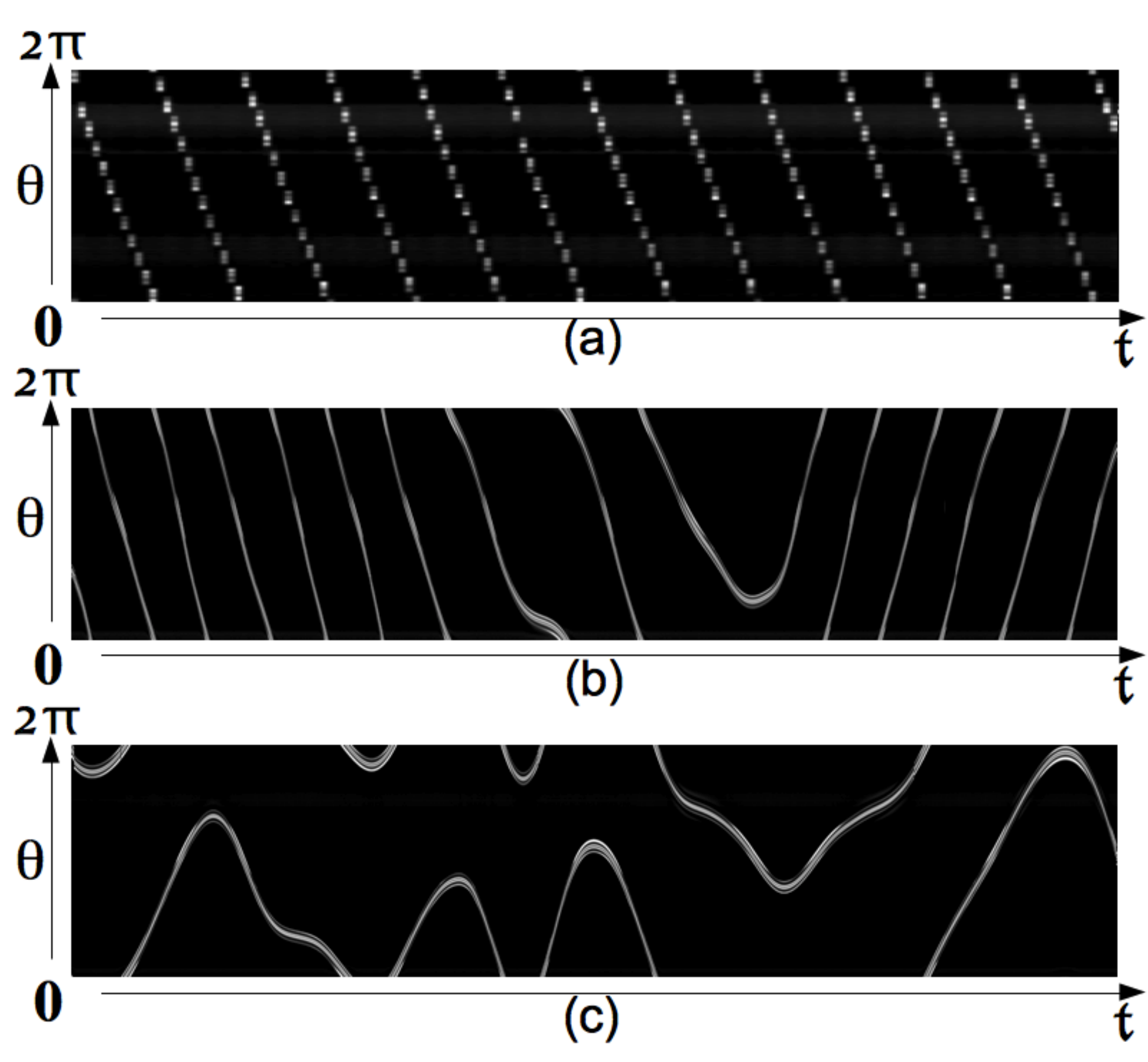}  \\
      \includegraphics[width=0.9 \columnwidth]{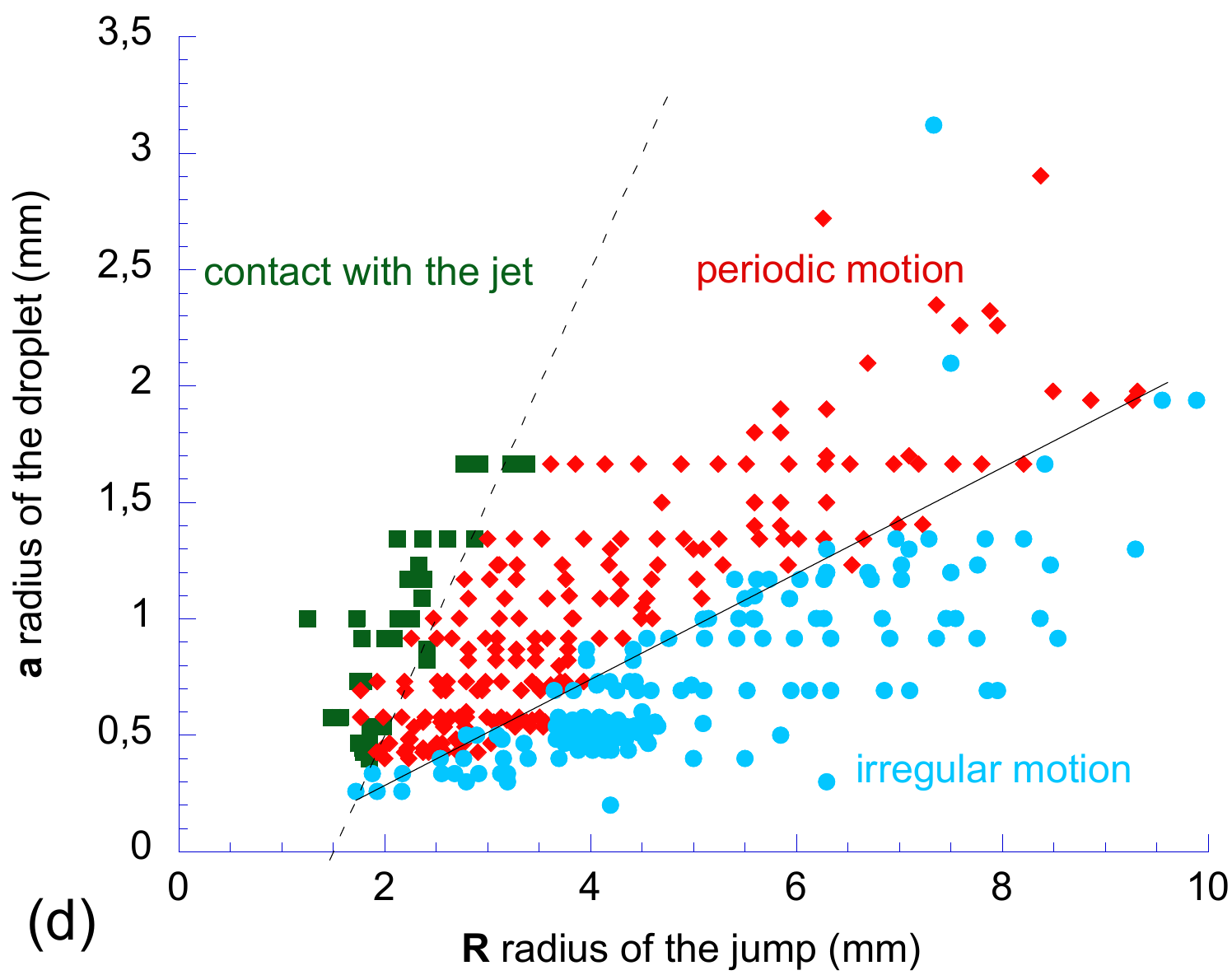}      
     \end{tabular}
    \caption{\label{fig2} Spatio-temporal diagrams obtained for a drop ($a = 1.1$ mm) in the periodic regime (a) ($R = 4.6$ mm and the record time is $t_{tot} = 7.4$ s) and in the irregular regime (b-c). Case (b) ($R = 5.5$ mm and $t_{tot} = 28.8$ s) displays a single reversal of the sense of rotation just above the transition between periodic and irregular motion, while case (c) ($R = 5.9$ mm and $t_{tot} = 22.4$ s) is obtained well above this one. (d) Phase diagram for the different kind of drop motions. ($\bullet$) irregular motion, ($\blacklozenge$) periodic motion and ($\blacksquare$) drop in contact with the jet}
\end{figure}
	 \indent A picture of the experiment is reproduced in Fig.~\ref{fig1}. A jet of silicone oil (viscosity $20$ cS, surface tension $20.6$ mN/m, density $0.95$) issued from a vertical tube of internal diameter $3$ mm, hits the center of a transparent glass disk placed $4$ cm below the outlet and of radius $R_0=15$ cm. With these boundary conditions (absence of a wall fixing the outer height of the jump) we observed that the studied hydraulic jump was of type I  \cite{Bohr_96} (i.e., unidirectional surface flow) and that the radius of the jump $R$ increased with the flow rate $Q$ while remaining very close to a power law $R =\alpha  Q^{\beta} $, where $\alpha =55 \pm 10 $ and $\beta =0.77 \pm 0.02$ (units used for Q and R are respectively here $m^3.s^{-1}$ and $m$). A constant level tank supplied with liquid by a gear pump is used in order to suppress any flow rate pulsation. Millimeter sized drops of the same fluid were deposited directly inside the jump and small enough ones remained trapped at the shock front \cite{Sreenivas_JFM}.\\
	\indent When a drop does not touch the impinging jet (i.e. when the distance between the jump and the jet is larger than the drop width) one observes two different drop dynamics: periodic and aperiodic. To characterize these phenomena, movies of the drop were recorded from below, through the glass plate. From the obtained frames the gray level evolution is extracted on a circle in order to obtain spatio-temporal diagrams giving the law $\theta(t)$ where $\theta$ is the angular position of the drop. Three examples are given in Fig.~\ref{fig2}:\\
	\indent (i) Fig.~\ref{fig2} (a) illustrates the periodic motion: the drop orbitates periodically along the jump with a frequency ranging between a few tenths of Hz and a few Hz. This frequency decreases when the jump radius increases.\\
	\indent (ii) Fig.~\ref{fig2} (c) illustrates the \enquote{irregular} motion: unlike the periodic one this regime is not characterized by a single orbital period. Speed variations are observed but also reversals of the sense of rotation. \\
	\indent (iii) In the irregular case, just beyond the transition between periodic and irregular motion, one observes sudden reversals of the drop rotation sense, separating sequences of quasi-regular motion in opposite directions (see Fig.~\ref{fig2} (b)).\\
	\indent In principle, when the geometry of the experiment is fixed (nozzle radius, impact distance,...) one only needs two parameters to characterize the different regimes : the drop radius $a$ and the jump radius $R$. A phase diagram is presented in Fig.~\ref{fig2} (d) where three different symbols are used for periodic and irregular motions, and for a drop in contact with the jet. In this last case one also observes complex motions (static, regular and irregular motions) but not well defined and with a lower drop velocity. As expected the dashed line separating the contact with the jet and the periodic motion is given by solving the equation $R=2a+R_J$ where $R_J$ is the radius of the jet that we find to be almost constant and equal to approximately 1.5 mm. The continuous line separating the periodic and aperiodic motions is empirical but can be fitted by the following law $R=\beta a+R_c$ where $\beta =4.4$ and $R_c=0.84$ mm.\\
	\indent A model of the drop motion can be built as follows \cite{Pirat_gyroscopic_droplet}. Entrained by the radial flow of the bath, the drop acquires an internal rotation with a kinetic momentum parallel to the shock front. If a perturbation shifts slightly the drop, the kinetic momentum conservation leads to the appearance of a radial component of rotation (See Fig.~\ref{fig4} (b)). In such a situation, an active torque should appear, amplifying the initial perturbation and leading to a self-sustained orbital motion. To recover this, one models the drop as a rigid sphere of radius $a$ with two contact points A and B. The situation is described in Fig.~\ref{fig4} (a).We assume that the exchanged forces at these points are simply viscous frictions through a sheared air film: $\overrightarrow {F_A}=\eta_a(S_A/d_A)(\overrightarrow U_A-\overrightarrow V_G-\overrightarrow\Omega \times \overrightarrow {GA})$ and $\overrightarrow {F_B}=\eta_a(S_B/d_B)(\overrightarrow U_B-\overrightarrow V_G-\overrightarrow\Omega \times \overrightarrow {GB})$, where $  \eta_a$ is the air dynamic viscosity and the quantities $S_A, S_B, d_A, d_B, \overrightarrow U_A, \overrightarrow U_B$ designate in this order : the contact surfaces, the local air layer thickness and the surface speed of the flow at the points A and B. G is the center of mass of the drop, $\overrightarrow V_G $ is the speed of the center of mass of the drop and $\overrightarrow\Omega$ designates  the angular velocity vector of the drop around G in the laboratory frame. \\
	The kinetic momentum conservation equations lead to :
	\begin{align}
& \frac {d\Omega_r}{dt}+\frac{1}{\tau_A}(\Omega_r+\frac{L}{a}\omega)  = \omega \Omega_\theta \notag \\ 
& \frac {d\Omega_\theta}{dt}+(\frac{1}{\tau_A}+\frac{1}{\tau_B})\Omega_\theta = -\omega \Omega_r- \frac{1}{a}(\frac{U_A}{\tau_A}+\frac{U_B}{\tau_B}) \notag \\
& \frac {d\Omega_z}{dt}+\frac{1}{\tau_B}(\Omega_z+\frac{L}{a}\omega) =0
	\end{align}
	 	\begin{figure}[t]
	    \centering
    \begin{tabular}{cc}
      \includegraphics[width=0.9 \columnwidth]{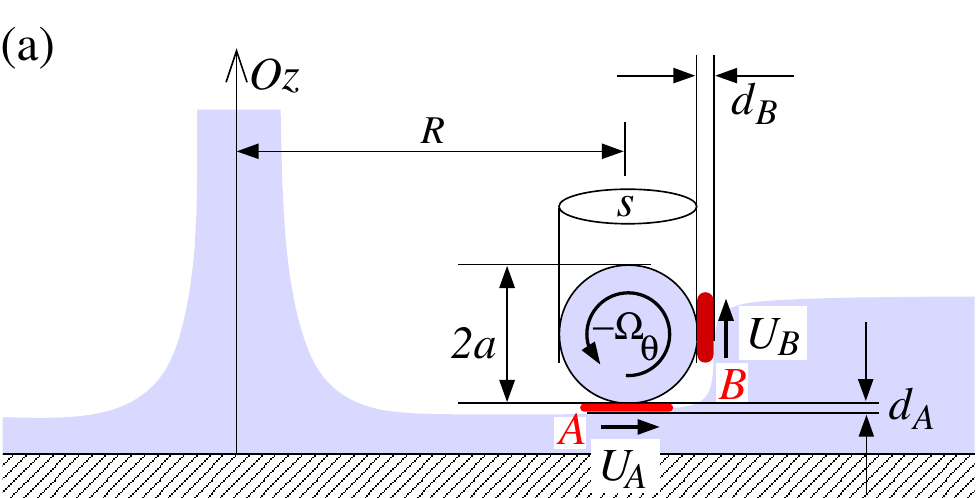}  \\
      \includegraphics[width=0.5 \columnwidth]{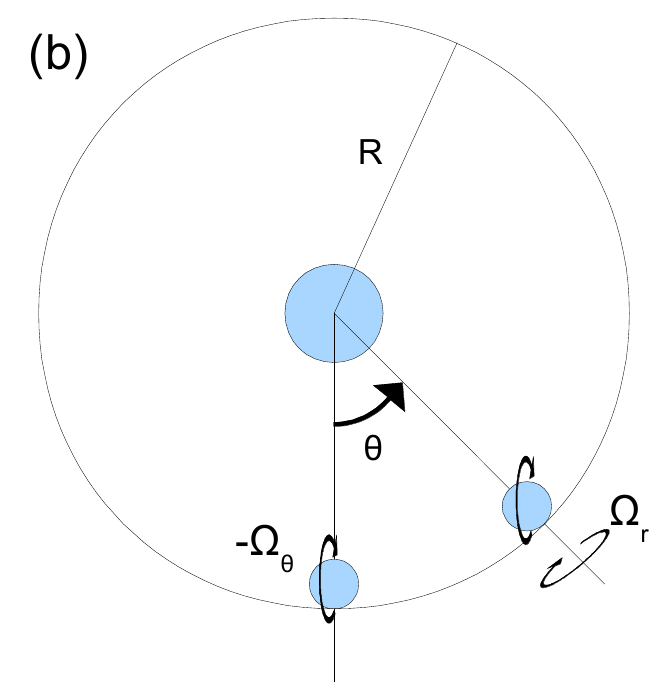}      
     \end{tabular}
    \caption{\label{fig4} (a) Notations and structure of the problem expected in a radial vertical plane containing the jump central axis. The drop is rotating very fast because of the shear stress transmitted across the air film. (b) Principle of the instability (top view). The drop tries to keep constant its kinetic momentum, which develops an active radial component of the rotation vector that tends to amplify the drop displacement.}
\end{figure}
	These equations must be coupled with the evolution equation for $\theta$ that can be deduced from the fundamental principles of dynamics :
	\begin{equation}
\frac {d\omega}{dt}+\frac{2}{5}(\frac{1}{\tau_A}+\frac{1}{\tau_B})\omega=-\frac{2}{5}\frac{a}{L}(\frac{\Omega_r}{\tau_A}+\frac{\Omega_z}{\tau_B})
\end{equation}
	Where  $\omega=\frac{d\theta}{dt}$ designates the orbital speed and $L=R-a$ is the radius of the orbit described by the drop. The two characteristic times are $\tau_A=\frac{8}{15}\pi a^3\frac{\rho_l}{\eta_a}\frac{d_A}{S_A}$ and $\tau_B=\frac{8}{15}\pi a^3\frac{\rho_l}{\eta_a}\frac{d_B}{S_B}$ where $\rho_l$ designates the mass density of liquid. 
	\indent There are only two stationary solutions to these equations. A first trivial one is:
		\begin{align}
 & \omega=\Omega_r=\Omega_z=0 \notag \\ 
 & \Omega_\theta=-\frac{1}{a}\frac{\frac{U_A}{\tau_A}+\frac{U_B}{\tau_B}}{\frac{1}{\tau_A}+\frac{1}{\tau_B}}
	\end{align}
	We have checked numerically that this solution is unstable, in accordance with our experimental observations. There is also a second solution defined by:
	\begin{align}
 & \Omega_\theta =0 \notag \\ 
 & \Omega_r=\Omega_z=-\omega \frac{L}{a} \notag \\
 & \omega^2 =\frac{U_A}{L\tau_A}+\frac{U_B}{L\tau_B}
 \end{align}
 		\begin{figure}[b]
\begin{centering}
\includegraphics[width=0.6 \columnwidth]{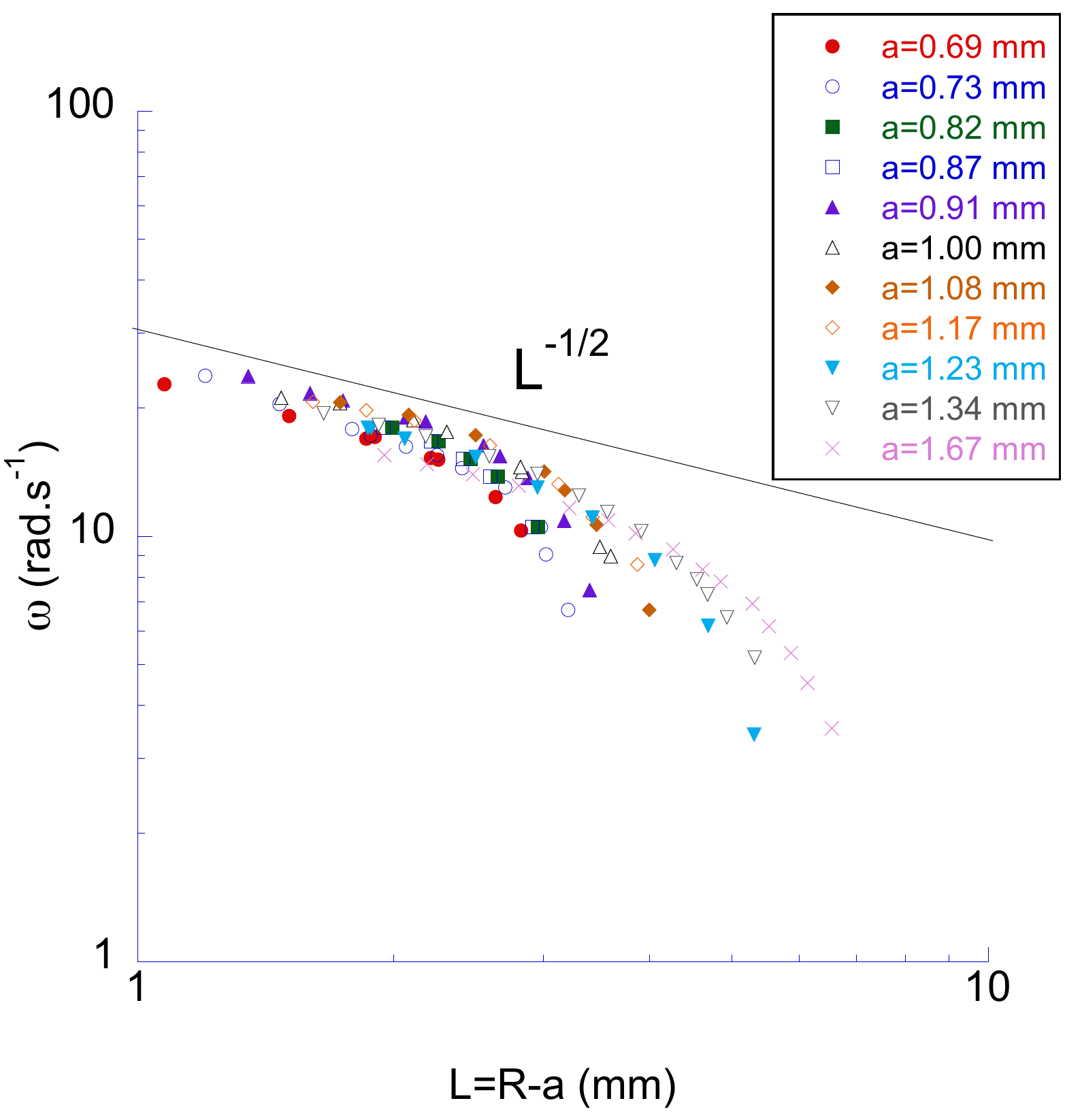}
\caption{\label{fig5}  Orbital speed of the drop $\omega=d\theta/dt$ versus the orbit radius followed by the drop: $L=R-a$. A large range of drops radii is reported: $0.69$ mm - $1.67$ mm }
\end{centering}
\end{figure}	
which corresponds to the periodic orbital state of our system. Complementary experiments showed that the velocity $U_A$ and  $U_B$  have almost the same magnitude (about $0.3\  m.s^{-1}$) and are nearly independent of the flow rate $Q$ (for a typical range of $5-60\ cm^3.s^{-1}$) . It seems also reasonable to suppose that $d_A$ and $d_B$, the local air layer thickness at the points A and B, are independent of the flow rate too (they could only be fixed by the surface speed and by the drop properties). So one can assume that $\tau_A$ and $\tau_B$ are independent of the flow rate.  We thus obtain that the orbital speed should scale as $\omega \propto L^{-1/2}$. This conjecture is tested in Fig.~\ref{fig5} by increasing the radius (through a modification of the flow rate) for different drop radii. The right behavior is obtained for low jump radii, but a decrease for large values of $L$ still needs to be explained. In addition, this model is unable to capture the aperiodic motion.\\
\indent The previous model is based on the hypothesis of an internal rotation of the drop but this phenomenon remains to be checked. We therefore performed some experiments in order to observe this rotation by injecting coal particles inside the drop. Pictures of this rotation are reproduced in Fig.~\ref{fig6}. The internal rotation is quite difficult to observe because of the drop motion inside the jump. To overcome this difficulty, we slightly inclined the jump plane (typically a few tenths of degrees), in order to maintain the drop globally motionless in the jump. Doing this, the solution (3) of equations (1) is selected, this solution being stabilized by the plate inclination \cite{Pirat_gyroscopic_droplet}. Assuming that the speeds are proportional (or even equal), which seems rather reasonable, one obtains the scaling law:
	 \begin{align}
\Omega_\theta=-\frac{1}{a}\frac{\frac{U_A}{\tau_A}+\frac{U_B}{\tau_B}}{\frac{1}{\tau_A}+\frac{1}{\tau_B}}\propto -\frac {U_A}{a}
	\end{align}	
		\begin{figure}[t]
\begin{centering}
\includegraphics[width=0.8 \columnwidth]{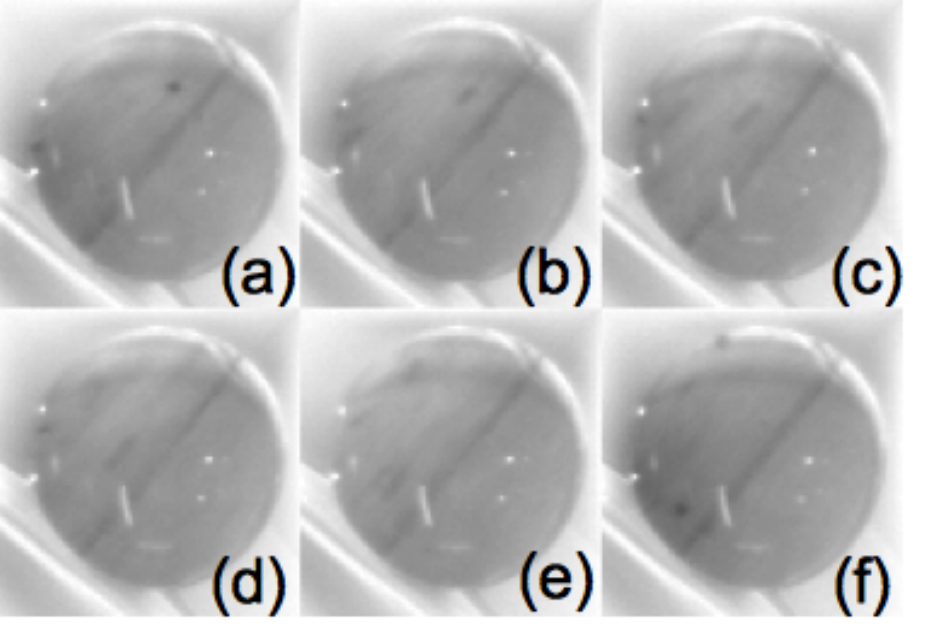}
\caption{\label{fig6} Coal particles inside a drop ($a=1.6$ mm and $R=1.3$ cm). Image sequence (the time between two successive pictures is about $10$ ms). The black points are coal particles.}
\end{centering}
\end{figure}
\begin{figure}[!t]
\begin{centering}
\includegraphics[width=0.8 \columnwidth]{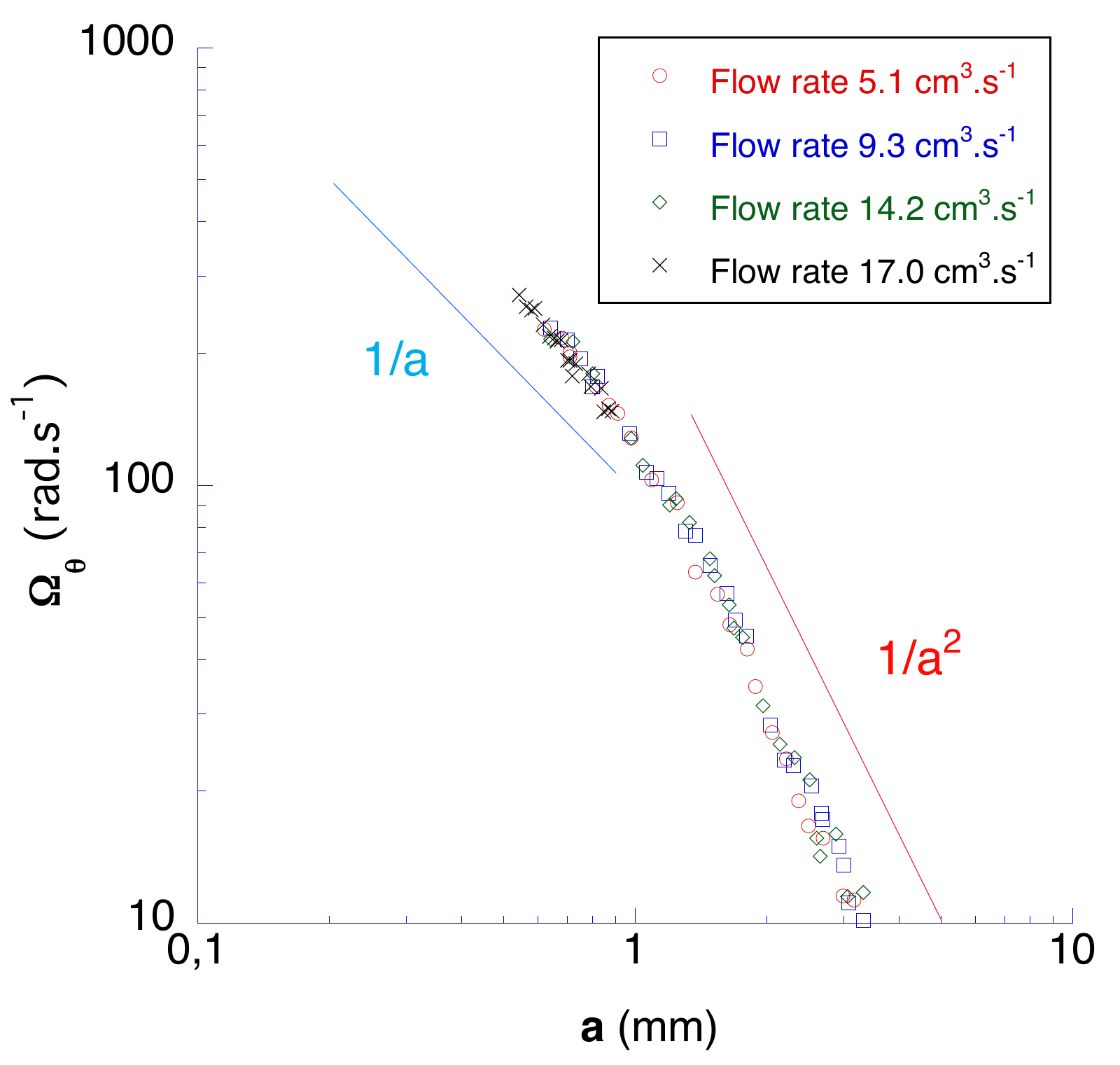}
\caption{\label{fig7} $\Omega_\theta$ versus drop radius observed for the stabilized inclined static case. Different flow rates are considered but all the curves are collapsing on a single master curve without any rescaling, which suggests that the surface velocity at the jump could be the same for each flow rate.}
\end{centering}
\end{figure}
	\indent Our experimental results, reproduced on Fig.~\ref{fig7}, highlighted a more complex situation : one can distinguish clearly two zone extremely different by increasing the drop radius:\\
	\indent (i) For \enquote{small} drops of radii $a < 1\ mm$ (remind that the capillary length $l_c$ for silicone oil is about $1\ mm$) , in the limit case of the smallest radii, we recover asymptotically the $\frac{1}{a} $ law. \\
	\indent (ii) For larger drops ($a>1\ mm$) one can observe a frequency law in $\Omega_\theta=U_A \frac{l}{a^2}$, where $l$ is about 0.4 mm.\\
	\indent Moreover, it is interesting to notice that all the curves presented here are collapsing on a single master curve without any rescaling. In other words the rotation frequency does not depend on flow rate. This is consistent with what we report above, i.e. that, in our specific situation (boundary conditions, flow rate range, viscosity...), $U_A$ and $U_B$ are nearly independent of flow rate. \\
	\indent The fact that the frequency law has a cross over around the capillary length suggests that the hypothesis of a rigid sphere on a sheared gas film is too rough and that the drop deformations have to be taken into account, with also possibly some local reversal of the air flow due to the lift applied on the drop. Indeed one can observe experimentally that the drops were slightly flattened by gravity. This is well known in the literature the drop is often modeled by a truncated sphere \cite{Gouttes_perles_deform}, or a truncated sphere sustained by a non axisymetric air pocket \cite{Neitzel_02}. \\ 
	\indent A new kind of \enquote{non wetting} dynamics for a drop in non coalescence state inside a horizontal hydraulic jump has been considered, with in particular two kinds of orbital motions: periodic and irregular. A model has been proposed and provides good agreement for the periodic orbital motion. An internal rotation of the drop has been highlighted and validates our theories. The frequency law illustrates the need for further investigations on this fascinating object. The structure of the hydraulic jump has also been studied with an original mean (introducing drop at the shock front) and reveals some unexpected observations about the local surface speed at the jump (does it depend or not upon flow rate ?) that would deserve further investigations. 
	\acknowledgements{Acknowledgements. We are grateful to Antoine Fruleux for fruitful discussions. The authors thank L. Rhea, and M. Receveur for technical assistance. This work was sponsored by the French National Agency for Research (ANR Freeflow).}
\bibliography{bibliojump.bib}
\end{document}